\documentstyle[aps,12pt,manuscript]{revtex}
\begin{document}
\preprint{INJE--TP--96--4}
\def\overlay#1#2{\setbox0=\hbox{#1}\setbox1=\hbox to \wd0{\hss #2\hss}#1%
\hskip -2\wd0\copy1}

\title{Propagating waves in an extremal black string}

\author{ H.W. Lee$^1$, Y.S. Myung$^1$, Jin Young Kim$^2$ and D.K. Park$^3$ }
\address{$^1$Department of Physics, Inje University, Kimhae 621--749, Korea\\
$^2$ Division of Basic Science, Dongseo University, Pusan 616--010, Korea \\
$^3$ Department of Physics, Kyungnam University, Masan 631--701, Korea} 

\maketitle

\vskip 1.5in

\begin{abstract}
We investigate the black string in the context of the string theories.
It is shown that the graviton is  the only  propagating mode in 
the (2+1)--dimensional extremal black string background. 
Both the dilation and axion turn out to be non-propagating modes.
\end{abstract}

\newpage
Recently there has been much progress in understanding the microscopic 
origin for
the black hole entropy. This was possible by using a new description of 
solitonic states in 
string theory\cite{Vafa,Callan,Horo1,Breck,John,Horo2}.  
For the simplest five-dimensional extremal black hole, Strominger and 
Vafa \cite{Vafa} counted the number 
of degeneracy corresponding to BPS-saturated states in the string 
theory for given charge.
And they showed that 
for large charge, the number of states increases as $e^{A/4}$, where $A$ 
is the area of the horizon for extremal
black hole. That is, the statistical interpretation for the 
Bekenstein-Hawking entropy 
was made possible in five dimensions. On the other hand, 
the five-dimensional black hole is just
a six-dimesional black string which winds around a compact internal 
circle\cite{Horo3}. 
The microstates of
five-dimensional extremal black hole arise from the fields moving 
around a circle in the internal
dimensions. In order to understand this situation, it is useful to take 
this internal direction as a space-time direction explicitly. This is a 
black string solution
in six dimensions.

As far as we know, full understanding of the black $p$-brane including the 
black string ($p=1$) 
was not completed, compared with the black holes. This is because the 
black $p$-brane solutions
are obtained in higher dimensions($D>4$)\cite{Gibb,Lars,Horo4}.
On the other hand, Horowitz and Tseytlin \cite{Tsey1,Tsey2} have done much work on finding
traveling wave solutions of string theories in various dimensions.
They related a limit of their $F$-model to the extremal black string 
in three dimensions.

The three-dimensional black string is used as a toy model for investigating 
higher dimensional black strings \cite{Lee}. Unlike the higher dimensional cases, 
the exact conformal field theory
is known for two and three dimensions. Thus the particle contents can 
also be determined.
A simple extension of Witten's construction for 
a gauged WZW model yields the three-dimensional charged black 
string\cite{Horn,Horo5}.  
This solution is characterized by three parameters: $M$ (mass), 
$Q$ (axion charge per
unit length), and $k$ (a constant related to the asymptotic value of 
the derivative of the dilaton).
For $0 < |Q| < M$, the black string is similar to the 
Reissner-Nordstr\"om black hole 
in four dimensions.  In addition to the event (outer) horizon 
($r_{EH} = M$), there exist
an inner horizon ($r_{IH} = Q^2/M$).  When $|Q| = M$, this corresponds to  
the extremal black string.
This is very similar to the five-dimensional extremal black string\cite{Gibb}. 

In this paper, we investigate the propagation of string fields in the 
$(2+1)$-dimensional
extremal black string background.  We follow the standard scheme for studying the black hole
physics\cite{Chan}. However there exists an apparent difference. For the black 
holes, the conventional method of metric perturbation ($h_{\mu\nu}$) is in such a way that 
the background symmetry ($\bar g_{\mu\nu}$) should be restored at the perturbation 
level. For example,  we take
$h_{\mu\nu}= h \bar g_{\mu\nu}$ in two-dimensional black hole\cite{Kim}.
In the study of Schwarzschild black hole, the spherical (background) symmetry
is also restored in the perturbation.
 As will be shown in (14), $h_{\mu\nu}$
takes the different form compared with the conventional one. Writing them in terms of
$v$ and $u$, $\bar g_{uv} \not=0,\bar g_{uu}=0$, 
whereas $h_{uv}=0, h_{uu} \not=0$.  Here $h_{\mu\nu}$ is introduced only for exploiting 
the null Killing symmetry
to study the extremal black string\cite{Garf1,Garf2}. 
Let us start with the $\sigma$-model action of  string theory\cite{Call,Rait}

\begin{equation}
S_\sigma = {-1 \over 4 \pi \alpha^\prime} \int d^2 \xi \sqrt{\gamma} 
\big (  \gamma^{ab} \partial_a X^\mu \partial_b X^\nu \tilde g_{\mu\nu} 
     +  \epsilon^{ab} \partial_a X^\mu \partial_b X^\nu B_{\mu\nu} 
     - {1 \over 2} \alpha^\prime R^{(2)} \Phi  \big ),
\end{equation}
where $R^{(2)}$ is the Ricci curvature of the world sheet.
As it stands, (1) is not conformally invariant because conformal invariance
is obviously broken by the dilaton ($\Phi$). It can be restored quantum 
mechanically by requiring that the $\beta$-function for 
$\tilde g_{\mu\nu}, B_{\mu\nu}$, and $\Phi$ vanish. To one loop, these  
equations are

\begin{equation}
{\tilde R}_{\mu\nu} - \nabla_\mu \nabla_\nu \Phi 
-{1\over 4} H_{\mu \rho \sigma} H_\nu^{~~\rho \sigma}  = 0,
\end{equation}

\begin{equation}
\nabla^2 \Phi + (\nabla \Phi)^2 - {4 \over k} - {1 \over 6} H^2   = 0,
\end{equation}

\begin{equation}
 \nabla_\mu (e^{\Phi} H^{\mu \nu \rho})  = 0,
\end{equation}
where $ H_{\mu \nu \rho} = 3 \partial_{[\mu} B_{\nu \rho]}$ is the axion 
corresponding to
$ B_{\nu \rho}$.
The above equations are also derived from the requirement that the 
fields must be an extremum
of the low-energy string action in the string frame\cite{Horn,Horo5,myung}

\begin{equation}
{\tilde S}_{l-e} = \int d^3 x \sqrt{-{\tilde g}} e^{\Phi}
\big \{ {\tilde R} + (\nabla \Phi)^2 + {4 \over k} - {1 \over 12} H^2 \big \},
\end{equation}
where the cosmological constant ${4 \over k}$ arises from the fact that 
the central charge is not
equal to the space-time dimensions. Here we set $\alpha'=2$ for simplicity.  
For the study of the black string, it is very useful to take the conformal
transformation  as\cite{Garf2} 
\begin{equation}
{\tilde g}_{\mu\nu} \equiv \Omega^2 g_{\mu\nu} = e^{-2 \Phi} g_{\mu\nu}.
\end{equation}
Then ond finds the  new action in the Einstein frame
\begin{equation}
S_{l-e} = \int d^3 x \sqrt{-g} 
  \big \{ R - (\nabla \Phi)^2 + {4 \over k} e^{-2 \Phi} - 
{1 \over 12} e^{4 \Phi} H^2 \big \}.
\end{equation}
The new equations of motion are given by
\begin{equation}
R_{\mu\nu} - \nabla_\mu \Phi \nabla_\nu \Phi 
-{1\over 4} e^{ 4 \Phi} H_{\mu \rho \sigma} H_\nu^{~~\rho \sigma} + 
{1 \over 6} g_{\mu\nu} 
e^{4 \Phi} H^2 + {4 \over k} g_{\mu\nu} e^{-2 \Phi} = 0,
\end{equation}

\begin{equation}
\nabla^2 \Phi  - {4 \over k}e^{-2 \Phi} - {1 \over 6} e^{4 \Phi} H^2   = 0,
\end{equation}

\begin{equation}
 \nabla_\mu (e^{4 \Phi} H^{\mu \nu \rho})  = 0.
\end{equation}
The static black string solution to the above equations is given by 
\begin{eqnarray}
&& \bar H_{rtx} = {Q \over r^2}, ~~~~~~\bar \Phi = \ln r - {1 \over 2} \ln k,
       ,   \nonumber   \\
&& \bar g_{\mu\nu} =
 \left(  \begin{array}{ccc} - {r^2 \over k}(1 - {M \over r}) & 0 & 0  \\
                             0 & {r^2 \over k}(1 - {N \over r}) & 0  \\
    0 & 0 & {1 \over 4} (1 - {M \over r})^{-1} (1 - {N \over r})^{-1} 
\end{array}   \right),
\end{eqnarray}
with $N \equiv Q^2/M $. 
Note that $\bar \Phi$ is a dimensionless quantity since $k$ has the 
dimension of length square. 
The above solution is clearly invariant under the translations
of both $t$ and $x$. For $r \to \infty$ and $k \to \infty$, the metric 
is asymptotically flat. 
Thus this represents a straight, static black string which is an extended object
with horizon.  When $Q=0$, $\bar H_{rtx}$ vanishes
 and this becomes a simple product of $dx^2$ and two-dimensional Witten's black hole.
 This is called as  a black string without charge and was discussed 
in\cite{Horn,myung}.
From now on we concentrate our study on the extremal limit of 
$N \to M(Q \to M)$.
In this limit, the line element is given by
\begin{equation}
ds^2= {r^2 \over k}( 1 -{ M \over r})( -dt^2 + dx^2 ) + 
{1 \over 4}( 1 -{ M \over r})^{-2}dr^2.
\end{equation}
Notice that this metric is not only static and translationally invariant, 
but boost invariant. 
This allows us to introduce two null-coordinates ($v= x +t, u = x -t$) and 
null Killing vector field ($ { \partial \over \partial v}$). Using these, 
(12) can be rewritten as
\begin{equation}
ds^2= {r^2 \over k}( 1 -{ M \over r}) du dv  + 
{1 \over 4}( 1 -{ M \over r})^{-2}dr^2.
\end{equation}
For our purpose, we introduce the  perturbation fields $({\cal H}, \phi, h)$ 
around the black string background  as\cite{Garf2,myung}
\begin{eqnarray}
&&H_{ruv} = \bar H_{ruv} + {\cal H}_{ruv}=\bar H_{ruv}(1+{\cal H}), 
\nonumber   \\   
&&\Phi = \bar \Phi + \phi,                      \nonumber   \\  
&&g_{\mu\nu} \equiv \bar g_{\mu\nu} + h_{\mu\nu}= 
\bar g_{\mu\nu} + h k_\mu k_\nu ,
\end{eqnarray}
where $\bar H_{ruv} = - M / {2 r^2}$ and $k_\mu$ is the null 
killing vector which satisfies
\begin{equation}
k^\mu k_\mu =0, ~~\nabla_{(\mu}k_{\nu)} =0, 
~~ k_{[\mu}\nabla_{\nu} k_{\rho ]} =0 .
\end{equation}
Here we choose $k_u = g_{uv}$ and $k^v=1$. For this metric perturbation, 
the harmonic gauge condition 
($\nabla_\mu h^\mu_{~\nu} = {1 \over 2} \nabla_\nu h^\mu_{~\mu}=0$) is trivially 
satisfied.
This ansatz for the metric perturbation is valid for the extremal black string in (12), 
but not for 
the black holes and nonextremal black string in (11). 
Conventionally the counting of gravitational degrees of freedom ($h_{\mu\nu}$) is as follows.
A symmetric traceless tensor has $D(D+1)/2 -1$ in $D$-dimensions.
$D$ of them are eliminated by the harmonic gauge condition. Also $D-1$ are eliminated by our freedom
to make further gauge transformations $\delta h_{\mu\nu}= \partial_\mu \xi_\nu
+ \partial_\nu \xi_\mu$ with $\partial_\mu \xi^\mu =0, \partial^2 \xi^\mu=0$.
Hence the number of drgrees of freedom is $D(D+1)/2 -1 - D -(D-1)=D(D-3)/2$.
We have no degrees of freedom in $D=3$. But we have one graviton gauge degrees of freedom($h$).
Thus it is clear that
 all gauge degrees of freedom of the graviton are not fixed in our ansatz (14).
From now on we follow the perturbation anaysis for the black 
holes. 
In order to obtain the  equations governing the perturbations, one has to 
linearize (8)-(10) as 
\begin{eqnarray}
&&  \delta R_{\mu\nu} (h) - 2 \bar \nabla_{( \mu} \bar \Phi \bar \nabla_{\nu )} \phi   
- e^{4 \bar \Phi } \{
 \bar H_{\mu \rho \sigma} \bar H_{\nu}^{~~\rho \sigma} \phi 
+{1 \over 2} (  \bar H_{\mu \rho \sigma} { \cal H}_{\nu}^{~~\rho \sigma}
- \bar H_{\mu \rho \sigma} \bar H_{\nu\alpha}^{~~~\sigma} h^{\rho \alpha} ) \}
                            \nonumber     \\
&& + { 1 \over 6} e^{ 4 \bar \Phi }  \{   \bar H^2 h_{\mu\nu}
 + \bar g_{\mu\nu} ( 4 \bar H^2 \phi + 2 \bar H_{\alpha \rho \sigma} 
{ \cal H}^{\alpha \rho \sigma}
  - 3 \bar H_{\kappa \rho \sigma} \bar H^{\alpha \rho \sigma} 
h^\kappa_{~\alpha} ) \} \nonumber \\ 
&& +{4 \over k} e^{ - 2 \bar \Phi } ( h_{\mu\nu} - 2 \bar g_{\mu\nu} \phi) = 0,
\end{eqnarray}

\begin{eqnarray}
&- h^{\mu\nu} \bar \nabla_\mu \bar \nabla_\nu \bar \Phi   
- \bar g^{\mu\nu} \delta \Gamma^\rho_{\mu\nu} (h) \partial_\rho \bar \Phi
+ \bar \nabla^2 \phi 
                            \nonumber     \\
& ~~~~ - { 1 \over 6} e^{ 4 \bar \Phi }  
  ( 4 \bar H^2 \phi + 2 \bar H_{\alpha \rho \sigma} 
{ \cal H}^{\alpha \rho \sigma}
- 3 \bar H_{\kappa \rho \sigma} \bar H^{\alpha \rho \sigma} h^\kappa_\alpha  )
+{8 \over k} e^{- 2 \bar \Phi } \phi = 0,
\end{eqnarray}

\begin{eqnarray}
 &\bar \nabla_\mu \{ e^{ 4 \bar \Phi }
                  (    {\cal H}^{\mu \nu \rho}
                     - \bar H_\alpha^{~~\nu\rho} h^{\alpha \mu}
                     - \bar H^{\mu~~\rho }_{~~\beta} h^{\beta \nu}             
                     - \bar H^{\mu \nu}_{~~~\gamma} h^{\gamma \rho} 
                     + 4 \bar H^{\mu \nu \rho} \phi )  \}  \nonumber \\
  &~~~~ - 4 e^{ 4 \bar \Phi } \bar H^{\alpha \nu \rho}     
     \delta \Gamma^\mu_{\mu \alpha} (h)  = 0, 
\end{eqnarray}
where  
\begin{eqnarray}
&\delta R_{\mu\nu} (h) = - {1 \over 2} 
(\bar \nabla_\mu \bar \nabla_\nu h^\rho_{~~\rho}
 +\bar \nabla^\rho \bar \nabla_\rho h_{\mu\nu}   
 -  \bar \nabla^\rho \bar \nabla_\nu h_{\rho\mu}   
 -  \bar \nabla^\rho \bar \nabla_\mu h_{\nu\rho}),   \\   
&\delta \Gamma^\rho_{\mu\nu} (h) = {1 \over 2} \bar g^{\rho\sigma} 
( \bar \nabla_\nu h_{\mu\sigma} + \bar \nabla_\mu h_{\nu\sigma} - 
\bar \nabla_\sigma h_{\mu\nu} ).
\end{eqnarray}
From (18) one can easily find 
\begin{equation}
{\cal H} = - 4 \phi.
\end{equation}
This means that on shell ${\cal H}_{ruv}$ is no longer an independent mode.  
The six equations from (16) are 
\begin{eqnarray}
(v,v) : & 0 = 0, \\
(v,r) : & \partial_v \phi =0, \\
(u,r) : & ( \partial_r + {{2 r -M} \over {r (r-M)}} ) \partial_v h = 
               { 4 k \over r^2(r-M)} \partial_u \phi, \\
(u,u) : & \partial^2_r h +{ 2 (3r -M) \over r(r-M)}\partial_r h +
             {{6r^2 -4 rM } \over r^2(r-M)^2} h = 0, \\
(u,v) : & \partial^2_v h + {16 (M^2 -r^2) \over r^4} \phi = 0, \\
(r,r) : & \partial_r \phi + {{r+M} \over r (r-M)} \phi = 0.
\end{eqnarray}
From (17) we obtain the dilaton equation
\begin{equation}
\bar \nabla^2 \phi - {8(2M^2 -r^2) \over r^4} \phi =0.
\end{equation}
Eq (22) is trivially satisfied.  Choosing $\partial_v h=0$, one 
finds $\partial_u \phi=0$.  It turns out 
that the only solution for the dilaton satisfying all 
eqs. (23), (24), (26), (27) and (28) is 
\begin{equation}
\phi={\cal H} = 0,
\end{equation}
which means that the dilaton and axion are the non-propagating modes in 
the black string background.
In order to solve (25), define $h$ as $h \equiv f(r) h'(r)U(u)$.  Then 
for $f(r)=1 /r(r-M)$,  plugging this into (25) leads to
\begin{equation}
\partial^2_r h' + { 2 \over {r-M}} \partial_r h'  = 0.
\end{equation}
The corresponding solution is given by
\begin{equation}
h'(r)= {1 \over r -M}.
\end{equation}
 
Now let us discuss our results. First of all, we compare our result 
with that of Garfinkle's case.  From Refs.\cite{Garf1,Garf2}, the relevant 
equation for (8) is
\begin{equation}
R^v_{~~u} 
-{1\over 4} e^{ 4 \Phi} H^v_{~~\rho \sigma} H_u^{~~\rho \sigma}  = 0.
\end{equation}
Each term of the above equation is identically zero in the black string 
background.
The linearized equation leads to
\begin{equation}
\bar g^{vu} (\delta R_{uu} +{ 1 \over 2} e^{4 \bar  \Phi}
\bar H_{u \rho \sigma} \bar H_{u \alpha}^{~~~\sigma} h^{\rho \alpha})
 - h( \bar R_{vu} -{1\over 4} e^{ 4 \bar \Phi} 
\bar H_{v \rho \sigma} \bar H_u^{~~\rho \sigma}) =0.
\end{equation}
This is reduced to (30), which can be rewritten as $\bar \nabla^2 h'=0$ with
$ \partial_v h'=0$.
This is just the Garfinkle's case.
Thus our solution is exactly the same as  Garfinkle's one.
Authors in \cite{Garf1,Garf2} used the generating technique to produce 
the solution (31) which represents  wave traveling in 
the black string background. 
In our case all informations to determine the solution of $h'(r)$ come
natually from the linearized equation (16). Thus our method is a standard one
to investigate the propagation of waves in the black string background.

The next is how to determine the form of $U(u)$. There is no additional 
constraint for determining $U(x,t)$ except $\partial_v U(x,t)=0$ within this
scheme.
 However we assume the normal mode solution of the form
\begin{equation}
U(x,t)= e^{-iEt} e^{-i\chi x}.
\end{equation}
From $\partial_v U(x,t)=0$ one finds $E= -\chi$. Then the form 
of $U(x,t)$ is given by
\begin{equation}
U(u)= e^{-iEu}.
\end{equation}
This is a plane wave along the $v$=constant.
And this is called the longitudinal wave as it carries only momentum $E$ 
along the string direction ($x$-direction) \cite{Horo4}. 
 Hence the graviton mode ($ h=f(r)h'(r)U(u)$) is the 
propagating wave in the black 
string background.

In conclusion, we find the solution which propagates in the 
three-dimensional extremal black string background. This corresponds to the 
graviton mode ($h$). Both the dilaton ($\phi$) and axion (${\cal H}$) are non-propagating modes. 
This seems to be a controversial result, when compared with the fact that
in three dimensions the graviton and axion are not propagating modes because
of gauge invariance. Explicitly, we consider the conventional counting of degrees of freedom.
The number of degrees of freedom for the gravitational field ($h_{\mu\nu}$) in 
$D$-dimensions is $(1/2) D (D -3)$ .  For $D=4$ Schwarzschild black
hole, we obtain two degrees of freedom. These correspond to Regge-Wheeler mode
 for odd-parity perturbation and Zerilli mode for even-parity perturbation \cite
{Chan}.  We have $-1$ for $D=2$. This means that in two dimensions
the contribution of graviton is equal and opposite to that of a spinless particle (dilaton).
In the 2d dilaton black hole, two graviton-dilaton modes are thus trivial gauge
artefacts \cite{Kim}.
For $D =3$, we have no  propagating 
graviton and the dilaton is a propagating one.
For example, it turned out that the dilaton and tachyon are two propagating
modes in the nonextremal black string\cite{myung}.
Similarly it can be conjectured that
there is no propagating graviton in the BTZ black hole\cite{Bana}.
This is because this black hole is equivalent (under T-duality) to the black string
solution\cite{Horn}. 
There does not exist the null Killing symmetry even for the extremal BTZ black hole.
It is emphasized that the conventional counting is
suitable for the black holes and nonextremal black string.
However, our model is the extremal
black string with the null Killing symmetry. 
In this case the graviton is progagating, while the dilaton is nonpropagating.
Note that the graviton may become a propagating mode by the transmutation of the 
degrees of freedom with other field (here, dilaton) in the extremal black
string background. This is similiar to the Higgs
mechanism for gauge fields in the Minkowski spacetime.
It seems that
the conventional counting for degrees of freedom is not valid for the extremal black string. 

\acknowledgments

This work was supported in part by the Basic Science Research Institute 
Program, Ministry of Education, Project NOs. BSRI--96--2441, BSRI--96--2413 
and by NONDIRECTED RESEARCH FUND, Korea Research Foundation, 1994.


\newpage

\begin{references}

\bibitem{Vafa} A. Strominger and C. Vafa, hep-th/9601029.
\bibitem{Callan} C. Callan and J. Maldacena, hep-th/9602043.
\bibitem{Horo1} G. Horowitz and A. Strominger, hep-th/9602051.
\bibitem{Breck} J. Breckenridge, R. Myers, A. Peet and C. Vafa, hep-th/9602065
\bibitem{John} C. Johnson, R. Khuri, and R. Myers, hep-th/9603061.
\bibitem{Horo2} G. Horowitz, D. Lowe, and J Maldacena, hep-th/9603195.
\bibitem{Horo3} G. Horowitz, gr-qc/9604051.
\bibitem{Gibb} G.W. Gibbons, G.T. Horowitz, Class. Quant. Grav. {\bf 12},
297 (1995).
\bibitem{Lars} F. Larsen and F. Wilczek, hep-th/9511064, hep-th/9604134.
\bibitem{Horo4} G. Horowitz  and D. Marolf, hep-th/9605224, hep-th/9606113.
\bibitem{Tsey1} G. Horowitz and A. Tseytlin,  Phys. Rev. {\bf D50}, 5204 (1994).
\bibitem{Tsey2} G. Horowitz and A. Tseytlin,  Phys. Rev. {\bf D51}, 2896 (1995).
\bibitem{Lee}  H. W. Lee, Y. S. Myung, J. Y. Kim, and D. K. Kim, hep-th/9610031.
\bibitem{Horn} J. Horne and G. Horowitz, Nucl. Phys. {\bf B368}, 444 (1992).
\bibitem{Horo5} G.T. Horowitz and D.L. Welch, Phys. Rev. Lett. {\bf 71}, 328 (1993).         
\bibitem{Chan} S. Chandrasekhar, {\it The Mathematical Theory of Black Hole}
                (Oxford Univ. Press, New York, 1983). 
\bibitem{Kim}  J.Y. Kim, H. W. Lee and Y. S. Myung, Phys. Lett. 
{\bf B328}, 291 (1994) ; Phys. Rev. {\bf D50}, 3942 (1994); 
Y. S. Myung, Phys. Lett. {\bf B334}, 29 (1994).
\bibitem{Garf1} D. Garfinkle and T. Vachaspati, Phys. Rev. {\bf D42}, 1960
                 (1990).                
\bibitem{Garf2} D. Garfinkle, Phys. Rev. {\bf D46}, 4286 (1992).                 
\bibitem{Call} C.G. Callan, D. Friedan, E,J. Martinec, and M.J. Perry, 
               Nucl. Phys. {\bf B262}, 593 (1985);
               T. Banks, Nucl. Phys. {\bf B361}, 166 (1991).
\bibitem{Rait} E. Raiten, Nucl. Phys. {\bf B416}, 881 (1994).          
\bibitem{myung} H. W. Lee, Y. S. Myung, and J. Y. Kim, Phys. Rev. 
{\bf D52}, 2214 (1995).
\bibitem{Bana} M. Banados, C. Teitelboim, and J. Zanelli, Phys. Rev. Lett. {\bf 69}, 1849 (1992).
\end{references}
\end{document}